# Expanding the Reach of Research Computing: A Landscape Study

Pathways Bringing Research Computing to Smaller Universities and Community Colleges

Dhruva K. Chakravorty

High Performance Research Computing, Texas A&M University, chakravorty@tamu.edu

SARAH K. JANES

Associate Vice Chancellor, San Jacinto College, sarah.janes@sjcd.edu

JAMES V. HOWELL

Associate Dean of Information Services, South Plains College, jhowell@southplainscollege.edu

LISA M. PEREZ

High Performance Research Computing, Texas A&M University, perez@tamu.edu

AMY SCHULTZ

Chief Relationship Officer, LEARN, amy.schultz@tx-learn.net

MARIE GOLDIE

Chief Relationship Officer, LEARN, mary.goldie@tx-learn.net

AUSTIN L. GAMBLE

Grants and Programs, LEARN, austin.gamble@tx-learn.net

RAJIV MALKAN

LoneStar College, Montgomery, TX, rajiv.r.malkan@lonestar.edu

HONGGAO LIU

High Performance Research Computing, Texas A&M University, honggao@tamu.edu

DANIEL MIRELES

High Performance Research Computing, Texas A&M University, danielmireles35@gmail.com

YUANQI JING

High Performance Research Computing, Texas A&M University, jingbec@tamu.edu

ZHENHUA HE

High Performance Research Computing, Texas A&M University, happidence1@tamu.edu

TIM COCKERILL

Texas Advanced Computing Center, The University of Texas at Austin, cockerill@tacc.utexas.edu

Research-computing continues to play an ever increasing role in academia. Access to computing resources, however, varies greatly between institutions. Sustaining the growing need for computing skills and access to advanced cyberinfrastructure requires that computing resources be available to students at all levels of scholarship, including community colleges. The National

Science Foundation-funded Building Research Innovation in Community Colleges (BRICCs) community set out to understand the challenges faced by administrators, researchers and faculty in building a sustainable research computing continuum that extends to smaller and two-year terminal degree granting institutions. BRICCs purpose is to address the technology gaps, and encourage the development of curriculum needed to grow a computationally proficient research workforce. Toward addressing these goals, we performed a landscape study that culminated with a community workshop. Here, we present our key findings from workshop discussions and identify next steps to be taken by BRICCs, funding agencies, and the broader cyberinfrastructure community.

## 1 INTRODUCTION

Cyberinfrastructure (CI) lies at the heart of today's data revolution. With digital literacy emerging as a "duty of citizenship," research computing is poised to play a progressively more important role in all areas of academics [1]. Sustaining this high level of success requires that computing resources be accessible to students at all levels of learning, especially at smaller universities and community colleges. Students at community colleges are now facing requirements to be able to expand their knowledge in an empirical way. Community college students whether vocational, technical, liberal arts, or in science will benefit from research skills. Simultaneously, research skills are essential to the success of students transferring from two-year colleges to four-year institutions. While these programs prepare students for the rigors of the new demands of digital literacy and data analytics, they need to form a more seamless transition both in curriculum development and implementation as well as specific technologies. Community colleges and research are not terms that are customarily thought of together, but as the importance of digital literacy emerges, this relationship must be strengthened.

Funded by the National Science Foundation CC* CIRA program, Building Research Innovations at Community Colleges (BRICCs, award number 2019136) [2] is a community-building group that explores ways to address the computing divide in institutions of higher education. At its core, BRICCs aims to identify pathways to bring innovations in research cyberinfrastructure to smaller institutions and community colleges. We have been aware of communities that are actively working in this space but have lacked a forum for them to share their work. There are several opportunities for collaboration between nascent research organizations and institutions proficient in research computing. Among its primary goals, BRICCs endeavors to increase awareness of available cyberinfrastructure resources within this community. The BRICCs program aims to create a platform where institutions can pool and share knowledge to ensure a seamless transition. Another goal is to collaborate and learn what community colleges are doing now and what they can be doing in the future in terms of incorporating research into their curriculum.

## 2 METHODS

Today, BRICCs is a collection of university administrators, Chief Information Officers (CIOs), faculty, students and research computing enthusiasts. The BRICCs community of volunteers, CIOs, and senior faculty members from smaller schools and institutions meet weekly via video conference calls. It is driven by the need to understand the Cyberinfrastructure needs and current challenges of the nascent research community. This requires that we identify the obstacles, and demonstrate the benefits of CI in research at the community college level. The community aims to identify strategies to incorporate research and CI into the curriculum, enabling access to research CI, and assist in developing articulation agreements between community colleges and terminal degree-granting institutions. To achieve these goals BRICCs offers to facilitate discussions between campus cyberinfrastructure groups, faculty, and administrators.

To understand the needs, challenges, and opportunities in the space, BRICCs leveraged the work done by CI collaborations such as the NSF CC* SWEETER CyberTeam [3], NSF Frontera [4], and CC* awardees to understand the needs and challenges faced by this community. Smaller schools and community colleges rarely have resources available for separate administrative computing facilities and research computing. Much like at several four-year institutions, the responsibility to develop computing services falls to the CIO. To understand the opportunities to grow computing in this space, BRICCs began the landscape study by conducting informal interviews of CIOs from community colleges, and university administrators from large and small schools in one-on-one and group settings. A set of common questions were developed to understand their CI challenges and elaborate on success stories. The collected information was correlated by interviews with groups offering CI services and advice in this space. With a view toward understanding the unforeseen challenges faced in implementing CI projects, an informal survey of smaller universities and community colleges with planned CI deployments was performed. BRICCs also engaged with the Texas Association of Community College CIOs (TACC-CIO) to further community engagement and advertise the BRICCs activities. From these engagements, we learned that there is interest by academic institutions to expand their cyberinfrastructure deployments and further research initiatives on campuses. BRICCs discovered that there are several groups working in this space and that connections need to be made.

"We are just trying to keep the lights on, I don't have time for anything new". This statement by one of the community college CIOs illustrates the fact that CIOs from smaller community colleges are time and budget constrained. This reality does not alleviate the fact that community college students need access to CI in order to frutifully participate in research, a prerequisite for many to further their education. Understanding this, we need to increase awareness of the available resources and the networks among the community. There are several opportunities to collaborate and learn from organizations and institutions proficient in research computing.

In light of this, BRICCs organized a workshop on October 18-19, 2021, to bring several groups working in this space to share information, and to highlight the resources available to smaller universities and community colleges. Since research depends on data analysis and digital literacy, it was determined that it would be appropriate to hold the seminar in conjunction with the Texas Association of Community Colleges-Chief Information Officer (TACC-CIO) annual meeting at South Plains College, a two-year institution in Levelland, Texas. This would allow for communication between faculty and CIOs about what infrastructure is in place or is needed for the incorporation of research into the curriculum.

With a view toward facilitating communications between these communities, the workshop adopted a structure that offered opportunities to learn about the technical, administrative, and academic aspects of this challenge over a period of two days. Various community mailing lists were leveraged to inform the community of this gathering. Offered in a hybrid in-person and virtual mode, the workshop was attended by over 110 participants from across the country. Participants included faculty, administrators, CIOs and IT professionals at two- and four-year institutions, along with representatives from industry and non-profits. To facilitate conversations between the participants and help foster new collaborations, virtual business cards that identified areas of interest were created for all registered participants. Sessions were scheduled in such a way to allow anyone from either the TACC-CIO group or the BRICCs group to participate in all sessions.

## 3 Results

The BRICCs workshop consisted of presentations, panels, and round-table discussions. Inspired in part by the structure of the NSF CC* program [6], the workshop focused on the coupled aspects of CI infrastructure adoption, and research enabled by CI. The first day of the workshop focused on technology for research and academics. CI projects of different scales that could assist smaller schools were represented in various ways. Speakers presented the opportunities for funding and organizations supporting CI. Thematic panels covered compliance, security, networking, and other CIO priorities. Lastly, discussions over lessons learned from implementations at smaller schools, and the challenges faced by CIOs during the COVID-19 era were discussed. The second day focused on the academics and research that use CI, including NSF Center for Occupational Research and Development Inc (CORD) [6] and Department of Labor Internships, research at Community Colleges, and Community Colleges student perspectives. Groups such as NSF CORD have completed a field analysis of industry needs and mechanisms to address them. The workshop showed that apprenticeship programs such as New Collar [7] are about developing skills and addressing the industry skills gap that we face in a world of fast-paced technology. We observed that a significant number of roles at companies like IBM don't require a traditional education or career path. What matters most are the skills and experiences to perform a role. The workshop concluded with a panel of Community Colleges student researchers talking about their passion for research and academia. The workshop's agenda with a list of speakers has been presented on the BRICCs website. Videos from the BRICCs landscape-workshop are available on the Texas A&M High Performance Research Computing (HPRC) channel on YouTube [8].

Following the workshop, BRICCs leadership held a series of meetings to discuss the workshop findings and identify recommendations for the future. Below are those findings, lessons learned, as well as recommendations to facilitate better research CI enablement at under-resourced schools and two-year institutions. Findings, lessons learned, and recommendations are addressed to CIOs, funding agencies, and the community. We also consider mechanisms to improve future workshops.

### 3.1 Workshop – Lessons Learned for BRICCs

A periodic convening of CIOs, program leaders, and faculty at small colleges and universities helps improve communications between them. While it helps convey the importance of their academic and professional pursuits, separate working convenings on specific projects or topics for the development of CI facilities and curriculum modules that will use them, is important. Future convenings should be either in-person or virtual but not both to ensure better interaction of participants. There is tremendous enthusiasm for computing and CI-enabled research at smaller schools. The traditional models offering computing via on-premise research computing centers, however, do not extend to these schools. We need to explore new models of offering computing resources for smaller schools and community colleges. These models could include a partnership model or expand to regional connectivity.

### 3.2 Workshop Findings

#### 3.2.1 *For the CI Community*

Research and academic programs at four-year and two-year institutions are geared and structured differently. The larger community is not aware of these differences. There is no documentation of learnings from CI implementations that is readily accessible to the community. Smaller institutions tend to learn about the challenges in engaging in research computing while integrating CI resources onto their campuses. Cybersecurity, compliance, and data-storage continue to remain issues of concern at institutions of all sizes.

#### 3.2.2 *For Funding-agencies and Foundations*

Small and/or rural colleges typically do not have the funds to support on-campus advanced cyberinfrastructure, both from an infrastructure acquisition and services perspective. CIOs in larger/urban settings may have more flexibility to adopt these practices. Federal programs are of considerable interest to both CIOs and faculty. Funds from the Coronavirus Aid, Relief, and Economic Security (CARES) act were available to several CIOs but usage varied across institutions. The NSF CC* and Major Research Instrumentation (MRI) programs offer other opportunities. Federally (CC*, Higher Education Emergency Relief Fund, CARE) available funds have alleviated some of the campus-networking challenges. Expertise is, however, necessary to manage these networks. Purchasing "network-consulting" support for higher education is an emerging opportunity.

### 3.2.3 *For CIOs*

Unlike their counterparts at larger and urban community colleges, CIOs working in smaller colleges and/or in rural settings do not have the funds to hire the number of technicians or the technicians with the level of skill to fully accommodate the technology support of the administration and faculty. In addition, CIOs at smaller and rural community colleges have both leadership and operational functions, limiting their ability to take on new opportunities. As such, institutional leadership has to work with CIOs to support activities that go past day-to-day functions and expand.

### 3.2.4 *For Academic Programs*

There is a vibrant research culture in preparing students for careers in computing at smaller institutions and community colleges. There is, however, limited curricular materials that leverage research computing at these institutions. Programs in cloud computing are being developed on several fronts, and could be leveraged to build curriculum using research computing. It is important to note that for community colleges, collaboration with various industries is needed to develop projects that truly relate to vocational technical problems. While community colleges would like to incorporate more research and data analytics into their curriculum, it is difficult to find the time to develop such projects.

## 3.3 Recommendations

### 3.3.1 *For BRICCs*

The BRICCs mission was appreciated by all who participated in the workshop. There is a need to create further awareness of the BRICCs community and the challenges small colleges and universities face related to advanced CI adoption. One way to do this would be to partner with groups such as the Eastern Regional Network and the regional CyberTeams. BRICCs could work with them to hold cross-cutting quarterly workshops open to statewide or regional participation. Workshops should focus on project development, faculty training on how to incorporate research projects into curriculum, and sharing of specific projects used for various industries. Another way to grow these communications would be to develop clusters for sub-group networking and sharing, and take opportunities to present at Texas Association for Career and Technology Educators (TACTE), and the Texas Community College Teachers and Administrators Association Conference (TCCTAA).

### 3.3.2 *For Funding Agencies*

While most community colleges have experience working with state agencies, support structures are needed to help faculty and CIOs submit grants to Federal funding agencies. Several community colleges do not have grants offices that actively support programs from the National Science Foundation. In some cases, the CIO themselves have to perform all the grant-related functions. Evaluation criteria for proposals from community colleges should be established and clarified by Federal agencies, in order to accurately represent the different research environment and needs at four-year and two-year institutions. Finally, equipment funding opportunities should also include support for day-to-day staffing.

### 3.3.3 *For the CI Community*

One of the critical findings was that even smaller-sized Institutions need to develop mechanisms to help CIOs better interact with their faculty, and communicate findings to the institution's administration. This could be in the form of steering committees, technical governance committees, community surveys and periodic mailers. Keeping in mind that the CIO offices at most rural and smaller institutions have limited human resources, we need to find ways to offer research computing resources to these schools while reducing the administrative models. As such, the community needs to explore new models of sharing computing between smaller institutions. Perhaps develop computing services in which collaborators share a resource. To facilitate cross-communication between researchers, faculty, administrations and CI personnel, we need to develop a standard vernacular. Terms such as "CI", "research computing", "networking" confused several attendees. There are opportunities to leverage trusted relationships between the Research and Educational Network providers (RENs) and flag-ship institutions for convening communities and workshops.

## 4 CONCLUSIONS AND FUTURE WORK

Stakeholders representing various communities and programs with overlapping interests in this space spoke about their programs and discussed foreseeable needs, and opportunities for collaboration. Perspectives from CIOs, faculty and students keen to perform CI-enabled research in these settings were collected. As community colleges understand the need for all levels within a workforce sector to have some digital skills including some ability to understand what data tells them, the colleges will need to offer modules in either IT courses or in workforce specific courses focused on these skills. Communication within colleges in which all workforce programs are involved will be paramount to developing these modules without redundancy but with more efficiency and effectiveness throughout the college programs. Those who attended the introductory BRICCs workshop were made aware of this basic need to establish channels of communication. In a simultaneous vein, we need to develop new models of academic research computing in which larger schools partner and support the under-resourced institutions. The recent NSF CC* solicitation highlights this approach in the regional computing track.

Following up on the recommendations and lessons learned from the workshop, Prof. Malkan, a BRICCs ambassador from Lone Star College, organized a session at the 2022 TCCTA conference in Dallas, TX. The session was attended by several faculty and senior members of Computer Science programs at two-year institutions. As a follow up, a workshop is being planned for August creating further awareness of the BRICCs grant and the challenges small colleges and universities face related to advanced cyberinfrastructure adoption. The leadership team will also hold quarterly meetings open to those who participated in the workshop and from those meetings develop clusters for the sub-group networking and sharing.

### ACKNOWLEDGMENTS

This work was supported by the National Science Foundation (NSF) award BRICCs - Building Innovation at Community Colleges (NSF #2019136). We thank staff at Texas A&M High Performance Research Computing, South Plains College and the TACC-CIO group for facilitating the BRICCs community workshop.### REFERENCES

[1] Dhruva K. Chakravorty, Lisa M. Perez, Honggao Liu, Braden Yosko, Keith Jackson, Dylan Rodriguez, Stuti H. Trivedi, Levi Jordan, and Shaina Le. 2021. Exploring Remote Learning Methods for User Training in Research Computing. The Journal of Computational Science Education 12, no. 2, 11-17. https://doi.org/10.22369/issn.2153-4136/12/2/2

[2] NSF BRICCs: https://www.nsf.gov/awardsearch/showAward?AWD_ID=2019136

[3] NSF CC* Team SWEETER: https://www.nsf.gov/awardsearch/showAward?AWD_ID=1925764

[4] NSF Frontera: https://www.nsf.gov/pubs/2020/nsf20018/nsf20018.jsp

[5] NSF CC* program: https://beta.nsf.gov/funding/opportunities/campus-cyberinfrastructure-cc

[6] NSF Center for Occupational Research and Development Inc (CORD): https://www.nsf.gov/awardsearch/showAward?AWD_ID=1501990

[7] The New Collar program: https://www.ibm.com/us-en/employment/newcollar/index.html

[8] Texas A&M High Performance Research Computing YouTube Channel: www.tinyurl.com/briccs6